\documentclass[prl,twocolumn,showpacs,preprintnumbers,amsmath,amssymb,superscriptaddress]{revtex4}

\usepackage{graphicx}
\usepackage{amsmath}
\usepackage{amsbsy}
\usepackage{dcolumn}
\usepackage{bm}
\begin{document}

\title{Decoherence processes of a quantum two-level system coupled to a fermionic environment}

\author{Naoyoshi Yamada}
\affiliation{Department of Applied Physics, Tohoku University,
Aoba 6-6-05, Aoba, Sendai 980-8579, Japan}
\author{Akimasa Sakuma}
\affiliation{Department of Applied Physics, Tohoku University,
Aoba 6-6-05, Aoba, Sendai 980-8579, Japan}
\affiliation{CREST, Japan Science and Technology Corporation (JST), Kawaguchi 332-0012, Japan}
\author{Hiroki Tsuchiura}
\affiliation{Department of Applied Physics, Tohoku University,
Aoba 6-6-05, Aoba, Sendai 980-8579, Japan}
\affiliation{CREST, Japan Science and Technology Corporation (JST), Kawaguchi 332-0012, Japan}

\date{\today}
\begin{abstract}

We study decoherence processes of an $S=1/2$ localized spin coupled to conduction band electrons
in a metal or a semiconductor via an Ising-like interaction.
We derive master equations for the density matrix of the localized spin, by tracing out all
degrees of freedom in the conduction electron system based on the linked-cluster-expansion technique.
%
%
It is found that the decoherence occurs more rapidly for the metallic case than for semiconducting case.
\end{abstract}

\pacs{ }

\maketitle


Because of its scalability and processability, a solid state device with localized spins
is considered as a promising candidate for quantum computing or spintronic devices \cite{zutic}.
Localized spins in a solid lose memory of their spin orientation through collisions with phonons, 
conduction electrons, and some other perturbations.
Thus to find a way to control the spin relaxation rate at least a few orders of magnitude is a key issue
in the field of quantum computation and spintronics.

In general, motion of localized spins in a conductor is damped by the transfer of angular momentum
and energy to the itinerant carriers \cite{tserAPL,tserRMP}.
Regarding the corresponding relaxation of a classical localized spin as a linear-response problem,
one of the authors (S.A.) has derived a Landau-Lifshitz-Gilbert type equation for the localized spin
based on the $s$-$d$ model in which the $s$-electron system is regarded as an environment weakly
coupled to the localized $d$-electron spin \cite{sakuma}.
He has shown that the relaxation function of the $s$-electron spins leads to the Gilbert type damping term
for the localized spin which corresponds to the retarded resistance function in a generalized Langevin equation.
The Ohmic form of the Gilbert term stems from the fact that the imaginary part of the spin susceptibility
of the itinerant electronic system is proportional to the frequency $\omega$ in low $\omega$ region.


Here we must consider the next question of the relaxation of a quantum localized spin ($S=1/2$) in a solid
interacting with its environment.
This is the problem of dissipation on quantum coherence, or the so-called problem of decoherence
of a quantum two-level system \cite{zurek,weiss}.
Quantum two-level systems interacting with an environment are fundamental models to describe
quantum relaxation or decoherence processes\cite{zurek}, and have been extensively investigated over
the past 20 years \cite{Hu,vladar,chang,shao,keil,virk}.
Although general theories of dissipative dynamics of a two-level system have been established,
none of these works directly discusses decoherence process of a quantum localized spin coupled to
conduction band electrons.

Thus in this paper, we study the dynamics of the $S=1/2$ localized spin interacting with a conduction electron system
in a metal or a semiconductor as a bath via an Ising-like interaction by using the kinetic equation of
the reduced density matrix.
We obtain the non-unitary master equation of the reduced density matrix for the spin by tracing out
the fermionic environment, following the standard decoherence calculations.
The decoherence factor of the spin is found to be linearly temperature dependent at a relatively
high temperature, while logarithmically dependent at low temperature region.
The effects of the band-filling of the conduction electrons on the decoherence factor will also be discussed.

The model we will consider here is so-called the Ising-Kondo model\cite{isingkondo} which consists of an $S=1/2$
localized spin coupled to a conduction electron system as an environment via an Ising-type interaction.
The Hamiltonian of this model is given as
\begin{eqnarray}
 {\cal H} &=& \frac{1}{2}\omega_{0}S_{z} + \sum_{k,\sigma}\varepsilon_{k}c_{k\sigma}^{\dagger}c_{k\sigma}
\nonumber \\
          &-& \frac{J}{N_{S}}\sum_{k,k'}S_{z}
               \left( c_{k\uparrow}^{\dagger}c_{k'\uparrow}
                    - c_{k\downarrow}^{\dagger}c_{k'\downarrow}\right) ,
\label{hamil}
\end{eqnarray}
where $\omega_{0}$ is the Larmor frequency, $c_{k\sigma}$ are annihilation operators for conduction electrons
with wavenumber ${\bm k}$ and spin $\sigma$, $N_{s}$ is the number of atoms in the system, 
and $S_{z}$ is the $z$ component of the localized spin.
It is worth noting that the coupling between the localized spin and the conduction electrons in this model
induces no spin flips, in other words, this model describes a phase decoherence only.
We assume that at the initial time $t=0$ the localized spin and the conduction electrons are decoupled,
and the conduction electrons are in thermal equilibrium.
Thus we can write the density matrix for the total system as
\begin{equation}
 \rho(0) = |\Psi_{0}\rangle\langle\Psi_{0}|\otimes w
\end{equation}
where $|\Psi_{0}\rangle$ is an initial quantum state of the localized spin
$|\Psi_{0}\rangle = \alpha|\uparrow\rangle + \beta|\downarrow\rangle$, 
and $w$ is the statistical operator for the conduction electron system given as
\begin{equation}
 w = \frac{e^{-\beta\sum_{k,\sigma}\varepsilon_{k}c_{k\sigma}^{\dagger}c_{k\sigma}}}
                 {{\mathrm{Tr}}\left(e^{-\beta\sum_{k,\sigma}\varepsilon_{k}c_{k\sigma}^{\dagger}c_{k\sigma}}\right)},
\end{equation}
where $\beta = 1/k_BT$.

Let us write the density matrix in the interaction representation
\begin{equation}
 \rho(t) = V(t)\rho(0)V(t)^{\dagger} ,
\end{equation}
with
\begin{equation}
 V(t) = T\exp\left( -\frac{i}{\hbar}\int_{0}^{t}dt'{\cal H}_{I}(t')\right) ,
\end{equation}
where $T$ symbolizes the time-ordering product and ${\cal H}_{I}$ represents the interaction term of the Hamiltonian,
that is, the third term in the right hand side of equation (\ref{hamil}).

The phase dynamics of the localized spin is described by the $2\times 2$ reduced density matrix $\tilde{\rho}(t)$
obtained by the operation of partial trace over the degrees of freedom of the conduction electrons:
\begin{equation}
 \tilde{\rho}(t) = {\mathrm{Tr}}_{\bm k}\left( V(t)|\Psi_{0}\rangle\langle\Psi_{0}|V(t)^{\dagger}\otimes w \right) .
\end{equation}

\begin{figure}
\includegraphics[width=7cm]{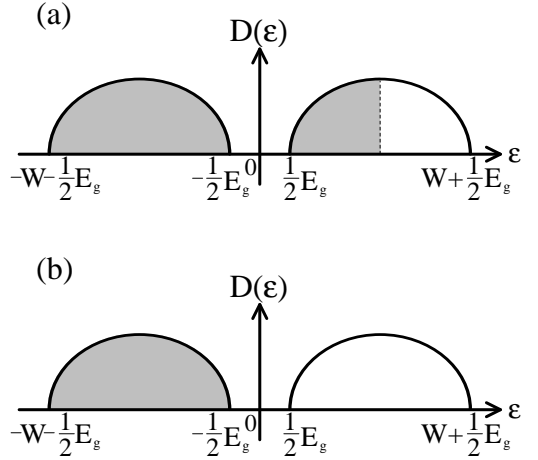}
\caption{
A schematic diagram of the density of states per atom $D(\varepsilon)$
for (a) a metal and (b) a semiconductor cases.
In the present calculation, $E_{g} = 0.01W$ and $E_{F} = (W+E_{g})/2$.
The hatched regions indicate the fermi sea at $T=0$.
}
\label{fig1}
\end{figure}

The diagonal terms of $\tilde{\rho}(t)$ are constant in time, i.e., $\tilde{\rho}_{\uparrow\uparrow}(t) = |\alpha|^{2}$
and $\tilde{\rho}_{\downarrow\downarrow}(t) = |\beta|^{2}$, because the interaction term in the present model
induces no spin flips.
The off-diagonal elements $\tilde{\rho}_{\uparrow\downarrow}(t)$ is given as
\begin{equation}
 \tilde{\rho}_{\uparrow\downarrow}(t) = \alpha\beta^{*}V(t)|\uparrow\rangle\langle\downarrow|V(t)^{\dagger}
\end{equation}
Within the linked-cluster-expansion technique, we obtain \cite{saikin,yamada}
\begin{equation}
 \tilde{\rho}_{\uparrow\downarrow}(t) = \alpha\beta^{*}\exp\left( -\Gamma(t) \right) ,
\end{equation}
with
\begin{eqnarray}
 \Gamma(t) &=& 4J^{2}\int_{-\infty}^{\infty}d\varepsilon_{1}\int_{-\infty}^{\infty}d\varepsilon_{2}
 \left( 1-f(\varepsilon_{1})\right) f(\varepsilon_{2})
\nonumber \\
 &&\times  D(\varepsilon_{1})D(\varepsilon_{2})
 \frac{ 1 - \cos\left(\frac{\varepsilon_{1}-\varepsilon_{2}}{\hbar}t\right) }{(\varepsilon_{1}-\varepsilon_{2})^{2}} ,
\end{eqnarray}
where $D(\varepsilon)$ is the density of states per atom.
To study the effects of the band-filling of the conduction electrons on the decoherence factor, 
here we use semi-elliptic forms of $D(\varepsilon)$, as shown in Fig. \ref{fig1}, with a normalization condition
$\int_{-\infty}^{\infty}d\varepsilon D(\varepsilon) = 2$.
We take the band gap as $E_{g} = 0.01W$,where $W$ is the band width.
For the metallic case (Fig. 1 (a)), the Fermi energy $E_{F}$ is taken to be at the center of the upper
band, i.e., $E_{F} = (W + E_{g})/2$, and for the semiconductor case (Fig. 1 (b)), $E_{F} = 0$.

\begin{figure}
\includegraphics[width=7cm]{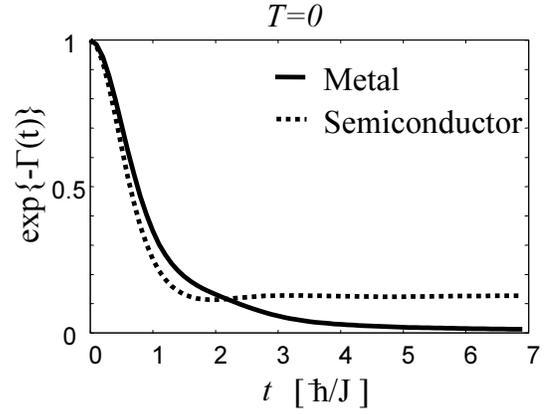}
\caption{
The time evolution of the off-diagonal element $\tilde{\rho}_{\uparrow\downarrow}(t)$ of the reduced density
matrix at $T=0$.
}
\label{fig2}
\end{figure}
\begin{figure}
\includegraphics[width=7cm]{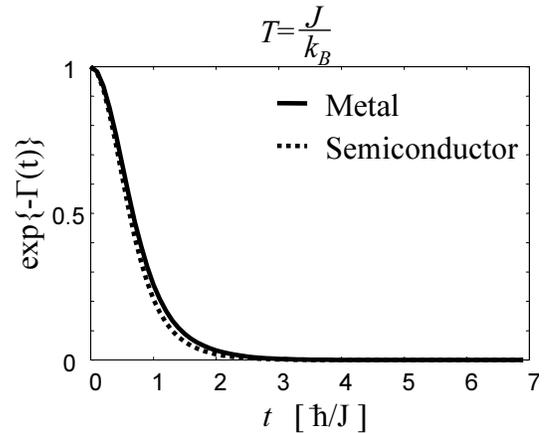}
\caption{
The time evolution of the off-diagonal element $\tilde{\rho}_{\uparrow\downarrow}(t)$ of the reduced density
matrix at $T=J/k_B$.
}
\label{fig3}
\end{figure}

Figure 2 shows the time evolution of the off-diagonal element $\tilde{\rho}_{\uparrow\downarrow}(t)$
of the reduced density matrix at $T=0$.
We have confirmed that the decoherence factor $\Gamma(t)$ is proportional to $t^{2}$ at $t \ll 1$.
As shown in Fig. 2, $\tilde{\rho}_{\uparrow\downarrow}(t)$ decays more rapidly for the metallic case
than that of the semiconductor case.
This is due to the low carrier density since the fermi energy is at the center of the band gap
in the semiconductor case.
Also unexpected behavior of the time evolution of $\tilde{\rho}_{\uparrow\downarrow}(t)$ can be seen in this figure,
that is, the dip-like structure of the decay curve around $t = 1$.
We speculate that this is a resonance effect due to the band gap, and detailed analysis of this
behavior will be reported elsewhere.

It is interesting to see the decoherence process in a relatively high temperature region.
We show the time evolution of $\tilde{\rho}_{\uparrow\downarrow}(t)$ at $k_BT = J$ in Fig. 3.
At this time,  $\tilde{\rho}_{\uparrow\downarrow}(t)$ decays exponentially for both the cases,
and the dip structure found in Fig. 2 cannot be seen because of the relatively high temperature than
$E_{g}/k_B$.

Finally, let us consider the decoherence process of the localized spin when it is magnetically interacting
with a bosonic environment, such as magnons.
We assume here that the magnons have a quadratic dispersion relation with a cutoff energy $\hbar\omega_{\rm m}$.
Following again the standard decoherence calculations, we can easily obtain an approximated form of the decoherence
factor $\Gamma(t)$; for example, $\Gamma(t)$ is roughly proportional to $t^{2}$ for $\omega_{\rm m}t\ll 1$,
and to $t^{3/2}$ for $\omega_{\rm m}t\gg 1$ at the low temperature limit.
As seen from figure 2, by contrast, $\Gamma(t)$ is roughly proportional to $t^{2}$ for $Jt/\hbar \ll 1$ and
to $\log t$ for $Jt/\hbar \gg 1$ at the zero temperature. 
Thus, to study the competition of the effects of conduction electrons and magnons on the decoherence process
is an interesting future problem\cite{yamada}.


In summary, we have derived the master equations for the localized $S=1/2$ spin (a two-level system) scattering
a bath of conduction band electrons in a metal and a semiconductor.
We have confirmed explicitly that the off-diagonal element of the reduced density matrix of the
localized spin decays more rapidly for the metallic case than that of the semiconductor case.

{\it Acknowledgments.} 

Numerical computation in this work was partially carried out at
the Yukawa Institute Computer Facility.
This work was partly supported by a Grant-in-Aid from the Ministry 
of Education, Science, Sports and Culture of Japan. 


\end{document}